\documentclass[aps,prb,twocolumn,superscriptaddress,showpacs]{revtex4}
\usepackage{graphicx}
\usepackage{latexsym}
\usepackage{amssymb}
\usepackage{amsmath}
\usepackage{amsfonts}
\usepackage{bm}
\usepackage{multirow}
\usepackage{color}

\newcommand{\beq}{\begin{equation}}
\newcommand{\eeq}{\end{equation}}
\newcommand{\beqn}{\begin{eqnarray}}
\newcommand{\eeqn}{\end{eqnarray}}

\begin{document}

\title{A Series of $(2+1)d$ Stable Self-Dual Interacting Conformal Field Theories}

%\author{Cenke's group}

%\author{Zhen Bi}

%\author{Yi-Zhuang You}

\author{Meng Cheng}

\affiliation{Department of Physics, Yale University, New Haven, CT
06520-8120, USA}

\author{Cenke Xu}

\affiliation{Department of Physics, University of California,
Santa Barbara, CA 93106, USA}

\date{\today}

\begin{abstract}

Using the duality between seemingly different $(2+1)d$ conformal
field theories (CFT) proposed recently
Ref.~\onlinecite{son,maxashvin,wangsenthil0,wangsenthil1,wangsenthil2,xudual,mross,karchtong,seiberg1,seiberg2},
we study a series of $(2+1)d$ stable self-dual interacting CFTs.
These CFTs can be realized (for instance) on the boundary of the
$3d$ bosonic topological insulator protected by U(1) and
time-reversal symmetry ($\mathcal{T}$), and they remain stable as
long as these symmetries are preserved. When realized as a
boundary system, these CFTs can be driven into anomalous
fractional quantum Hall states once $\mathcal{T}$ is broken. We
demonstrate that the newly proposed dualities allow us to study
these CFTs quantitatively through a controlled calculation,
without relying on a large flavor number of matter fields.

\end{abstract}

\pacs{}

\maketitle

\section{Introduction}

%{\it --- Introduction}

Analytical studies on interacting $(2+1)d$ conformal field
theories (CFT) usually rely on a large flavor number of matter
fields, unless the theory has supersymmetry. For instance, the
$(2+1)d$ quantum electrodynamics (QED) with a large flavor number
of massless Dirac fermions is a stable CFT if all gauge-invariant
fermion bilinear operators are forbidden by symmetry. The usual
wisdom is that, this CFT can be studied reliably through a $1/N$
expansion ($N$ is the number of Dirac fermions), as long as $N$ is
larger than some critical number. When $N$ is small, not only is
the $1/N$ expansion no longer reliable, this CFT could be unstable
against spontaneous mass generation~\cite{chiral}.

Recent studies on the bulk duality between gauged topological
insulators~\cite{wangsenthil0,wangsenthil1,wangsenthil2,maxashvin}
have led to a conjectured duality between a single noninteracting
massless $(2+1)d$ Dirac fermion and a $(2+1)d$ QED with one flavor
($N=1$) of massless Dirac fermion~\footnote{For simplicity and
convenience, we will use the formalism developed in
Ref.~\onlinecite{son,wangsenthil0,wangsenthil1,wangsenthil2,maxashvin},
instead of the version in Ref~\onlinecite{seiberg1} which does not
require a special quantization of the flux of $a_\mu$.}: \beqn &&
\mathcal{L} = \bar{\chi} \gamma_\mu (\partial_\mu - i A_\mu) \chi
\cr\cr &\leftrightarrow& \bar{\psi} \gamma_\mu (\partial_\mu - i
a_\mu) \psi + \frac{1}{e^2} f_{\mu\nu}^2 + \frac{i}{4\pi}
\epsilon_{\mu\nu\rho}a_\mu
\partial_\nu A_\rho. \label{dual1} \eeqn The Lagrangian in the second
line above was also proposed earlier as a dual description of the
half-filled Landau level with a particle-hole symmetry~\cite{son}.
The symbol ``$\leftrightarrow$" stands for ``dual to". In this
equation, $a_\mu$ is a dynamical noncompact U(1) gauge field, and
$A_\mu$ is an external background U(1) gauge field. The ``flux
current" of $a_\mu$ is dual to the fermion current of $\chi$:
$\bar{\chi}\gamma_\mu\chi = \frac{1}{4\pi}
\epsilon_{\mu\nu\rho}\partial_\nu a_\rho $. In the context of
topological insulator, the physical meaning of this duality
mapping is that, the $4\pi$ flux of $a_\mu$, which is bound with
the strength-4 vortex of the ``Fu-Kane"
superconductor~\cite{fukane2} of $\psi$, is the fermion $\chi$;
and the fermion $\psi$ can also be viewed as the strength-4
vortex of the Fu-Kane superconductor of $\chi$. %There are two
%gauge fields in Eq.~\ref{dual1},

The strongest interpretation of this duality is that, the $(2+1)d$
QED with $N=1$ and weak coupling constant $e$ in the ultraviolet
will flow to a strongly interacting CFT in the infrared under
renormalization group, and this is the same CFT as a free Dirac
fermion under duality transformation. Eq.~\eqref{dual1} is the
fermionic version of the well-known boson-vortex
duality~\cite{halperindual,leedual}, which states that the O(2)
Wilson-Fisher fixed point is dual to the Higgs transition of the
bosonic QED with $N=1$ in $(2+1)d$. Recently it was shown that
Eq.~\eqref{dual1} is one example of a bigger ``web" of
dualities~\cite{seiberg1,seiberg2}.

This new duality sheds light on our understanding of CFTs with a
small flavor number of matter fields. Ref.~\onlinecite{xudual}
showed that the $(2+1)d$ QED with $N=2$ flavors is self-dual. This
is a fermionic version of the self-duality of the easy-plane
noncompact CP$^{1}$ model (a $N=2$ bosonic
QED)~\cite{ashvinlesik,deconfine1,deconfine2}. The self-duality of
the $N=2$ QED was also verified with different
derivations~\cite{karchtong,seiberg2}. Unlike the previous case
with $N=1$, there is no equivalent noninteracting
description of the $(2+1)d$ QED with $N=2$. %Since both sides of the
%duality have a SU(2) global flavor symmetry, the actual symmetry
%of this CFT is SO(4). This enlarged symmetry was first discussed
%in Ref.~\onlinecite{senthilfisher}.
Recent numerical studies on the $N=2$ QED indicates that this
theory could indeed be a scale-invariant CFT in the infrared
limit~\cite{qedcft} (although earlier study suggests a spontaneous
mass generation~\cite{kogut2}). However, it is difficult to study
the $N=2$ QED quantitatively using analytical methods because it
is unclear whether the standard $1/N$ expansion actually provides
useful information for $N=2$.

In this paper we study a series of self-dual QEDs with flavor
number $N=2$. The Lagrangian of these QED reads
\beqn \mathcal{L}
&=& \bar{\psi}_1 \gamma_\mu (\partial_\mu - i k a_\mu - i 2 n_B
B_\mu) \psi_1 + \bar{\psi}_2 \gamma_\mu (\partial_\mu - i a_\mu)
\psi_2 \cr\cr &+& \frac{i n_A }{2\pi} \epsilon_{\mu\nu\rho}a_\mu
\partial_\nu A_\rho + \cdots \label{qed1} \eeqn
We take the
convention that $\gamma_0 = \sigma^y$, $\gamma_1 = \sigma^z$,
$\gamma_2 = \sigma^x$. Again, $a_\mu$ is a dynamical U(1) gauge
field, while $A_\mu$ and $B_\mu$ are two external background U(1)
gauge fields. The fermion $\psi_1$ carries gauge charge $k$ of
$a_\mu$, and charge $2n_B$ of $B_{\mu}$; the fermion $\psi_2$
carries charge $1$ of $a_\mu$. Most importantly, the $2\pi-$flux
of $a_\mu$ carries charge $n_A$ of $A_\mu$, hence it is a
noncompact U(1) gauge field when $n_A\neq 0$.

The constants $k$, $n_A$ and $n_B$ in Eq.~\eqref{qed1} depend on
the physical realization of the theory. In section III we will
show that the case with an odd integer $k$ has a natural
realization as the boundary of a $3d$ bosonic SPT state, more
precisely it is the boundary of a bosonic topological insulator
with U(1) and time-reversal symmetry ($\mathcal{T}$). In this
paper we will mostly focus on the case with odd integer $k$. The
case with even integer $k$ will also be briefly discussed in
section III.

%The special case $k=1$ was studied
%in Ref.~\onlinecite{senthilfisher,xudual,seiberg2}.
%When $k > 1$, Eq.~\ref{qed1} no longer has a SO(4) flavor
%symmetry.

The theories in Eq.~\eqref{qed1} parameterized by $k$ are
interacting theories with no free theory dual. However, we will
show that the newly proposed dualities mentioned above can help us
study this CFT quantitatively with a $1/k$ expansion.

\section{Self-dual Conformal Field Theory}

%{\it --- Self-dual Conformal Field Theory}
\subsection{The cases with $k > 1$}

We now demonstrate that Eq.~\eqref{qed1} is self-dual for
arbitrary odd integer $k$. Following the duality in
Eq.~\eqref{dual1}, Eq.~\eqref{qed1} is dual to the following
theory: \beqn \mathcal{L} &=& \bar{\chi}_1 \gamma_\mu
(\partial_\mu - i b_\mu) \chi_1 + \bar{\chi}_2 \gamma_\mu
(\partial_\mu - i c_\mu) \chi_2 \cr\cr &+& \frac{i}{4\pi}
\epsilon_{\mu\nu\rho}a_\mu
\partial_\nu (k b_\rho + c_\rho + 2 n_A A_\rho) +
\frac{in_B}{2\pi} \epsilon_{\mu\nu\rho} b_\mu \partial_\nu B_\rho
\cdots \label{qed3}\eeqn In the derivation above we have used the
duality mapping that the fermion current $\bar{\psi}_1 \gamma_\mu
\psi_1$ is mapped to $\frac{1}{4\pi} \epsilon_{\mu\nu\rho}
\partial_\nu b_\rho$, and $\bar{\psi}_2 \gamma_\mu \psi_2 = \frac{1}{4\pi}
\epsilon_{\mu\nu\rho} \partial_\nu c_\rho$. Integrating out
$a_\mu$ will impose the following constraint: \beqn c_\mu = - k
b_\mu - 2 n_A A_\mu. \eeqn Thus the dual theory of
Eq.~\eqref{qed1} reads \beqn \mathcal{L} &=& \bar{\chi}_1
\gamma_\mu (\partial_\mu - i b_\mu) \chi_1 + \bar{\chi}_2
\gamma_\mu (\partial_\mu - i k b_\mu - i 2 n_A A_\mu ) \chi_2
\cr\cr &+& \frac{in_B}{2\pi} \epsilon_{\mu\nu\rho} b_\mu
\partial_\nu B_\rho. \label{qed2} \eeqn In the last equation above
we have performed a particle-hole transformation on the dual Dirac
fermion $\chi_2$. This theory Eq.~\eqref{qed2} takes exactly the
same form as Eq.~\eqref{qed1}, except that the gauge charges of
the two flavors of fermions are exchanged, and the roles of the
two external gauge fields $A_\mu$ and $B_\mu$ are also exchanged.
(One potential subtlety of the derivation above is the flux
quantization of the three dynamical gauge fields $a_\mu$, $b_\mu$
and $c_\mu$, which will be clarified in the next section when we
discuss the physical construction of the theory.)

Besides the two U(1) global symmetries, we can also impose
discrete symmetries to exclude the Dirac fermion mass terms. For
example, we can define the time-reversal transformation for
$\psi_j$ and $\chi_j$: \beqn \mathcal{T} : \psi_j \rightarrow i
\sigma^y \psi^\dagger_j, \ \ \ \chi_j \rightarrow i \sigma^y
\chi_j. \eeqn We can view the duality transformation as a $Z_2$
transformation, and the CFT under study also has this $Z_2$
self-dual symmetry. This is analogous to the ``mirror symmetry" of
the $(2+1)d$ supersymmetric field theories~\cite{mirror1,mirror3},
which also acts on the theories as a duality transformation.

We can rescale $a_\mu$, and define $\tilde{a}_\mu = k a_\mu$. Then
$\psi_1$ carries charge $1$ under $\tilde{a}_\mu$, and $\psi_2$
carries charge $1/k$. Thus with large $k$, $\psi_2$ is effectively
decoupled from the gauge field. Another way to understand this
statement is that, when $k \gg 1$, the dressed propagator of
$a_\mu$ is at order of $1/k^2$, thus the effect of $a_\mu$ on
$\psi_2$ is strongly suppressed with large $k$. For the same
reason, with large $k$, the dual fermion $\chi_1$ is effectively
decoupled from the dual gauge field.

The self-duality of Eq.~\eqref{qed1} gives us very helpful
quantitative information about the theory, at least for the
purpose of a $1/k$ expansion. For example, to compute physical
quantities to the first order of the $1/k$ expansion, we need to
know the gauge field propagator in the large$-k$ limit. In this
limit, because $\psi_2$ decouples from the dynamical gauge field,
in order to calculate the fully-dressed $a_\mu$ propagator one can
ignore $\psi_2$. Now the theory reduces to a QED with $N=1$, which
is dual to a single Dirac fermion $\chi$. The duality states that
the dual fermion current $J^{\chi}_{\mu} = \bar{\chi} \gamma_\mu
\chi = \frac{1}{4\pi} \epsilon_{\mu\nu\rho}\partial_\nu
\tilde{a}_\mu = \frac{k}{4\pi} \epsilon_{\mu\nu\rho}\partial_\nu
a_\mu$. In this limit the correlation function of $J^{\chi}_\mu$
can be computed exactly because $\chi$ decouples from any gauge
field: \beqn \langle J^{\chi}_\mu(p) \ J^{\chi}_\nu(-p) \rangle =
\frac{1}{16} |p| \left( \delta_{\mu\nu} - \frac{p_\mu p_\nu}{p^2}
\right). \eeqn This implies that the full $a_\mu$ propagator in
the large$-k$ limit reads \beqn G^a_{\mu\nu}(\vec{p}) =
\frac{\pi^2}{k^2 |p|} \left( \delta_{\mu\nu} - \frac{p_\mu
p_\nu}{p^2} \right). \label{propagator} \eeqn We have taken the
Landau gauge. The propagator of $b_\mu$ in the large$-k$ limit
takes the same form.

There is a slightly different way of deriving the propagator of
$a_\mu$: by varying with $A_\mu$ on both Eq.~\eqref{qed1} and
Eq.~\eqref{qed2}, we can conclude that $2\bar{\chi}_2 \gamma_\mu
\chi_2 = \frac{1}{2\pi} \epsilon_{\mu\nu\rho}\partial_\nu a_\rho$.
Then varying $b_\mu$ leads to the constraint $k\bar{\chi}_2
\gamma_\mu \chi_2 = - \bar{\chi}_1 \gamma_\mu \chi_1$. Since
$\chi_1$ decouples from the gauge field in the large$-k$ limit,
the $a_\mu$ propagator can be computed through the correlation of
the fermion current $\bar{\chi}_1 \gamma_\mu \chi_1$, and the
result will be the same as Eq.~\eqref{propagator}.

According to Ref.~\onlinecite{fisher1,fisher2}, a $(2+1)d$ CFT
with a U(1) global symmetry should have a universal conductivity.
Due to the last term of Eq.~\eqref{qed1}, a $2\pi-$flux of $a_\mu$
carries global U(1) charge $n_A$ of the external gauge field
$A_\mu$. Hence the global U(1) charge current of $A_\mu$ is
$J^A_\mu = \frac{n_A}{2\pi} \epsilon_{\mu\nu\rho}\partial_\nu
a_\rho$. We can use the propagator of $a_\mu$ to compute the
universal conductivity of the global U(1) charge transport: \beqn
\langle J^A_{\mu}(p) \ J^A_{\nu}(-p) \rangle = \tilde{\sigma}_A
|p| \left( \delta_{\mu\nu} - \frac{p_\mu p_\nu}{p^2} \right),
\label{universal}\eeqn where $\tilde{\sigma}_A$ is the universal
conductivity. Comparing Eq.~\eqref{universal} and
Eq.~\eqref{propagator} leads to the conclusion that with large $k$
the leading order universal conductivity is \beqn \tilde{\sigma}_A
= \frac{n_A^2}{4k^2}. \eeqn For the same reason, the universal
conductivity of the current of $B_\mu$ is $ \tilde{\sigma}_B =
\frac{n_B^2}{4k^2}$

The self-duality also identifies local operators on the two sides
of the duality. For example we can identify $\bar{\psi}_1\psi_1
\leftrightarrow - \bar{\chi}_1\chi_1$, $\bar{\psi}_2\psi_2
\leftrightarrow - \bar{\chi}_2\chi_2$. All these operators are odd
under time-reversal. This identification of operators will be
further clarified in section III. Using the fully-dressed gauge
field propagator, we can compute the scaling dimension of these
fermion mass operators at the leading order of $1/k$ expansion:
\beqn \Delta[\bar{\psi}_j \psi_j] = \Delta[\bar{\chi}_j \chi_j] =
2 - \frac{4}{3k^2}. \label{scaling}\eeqn The $1/k$ correction
comes from the fermion wave function renormalization and vertex
correction due to the coupling to the gauge field, like the
calculation in (for example) Ref.~\onlinecite{hermele2005}. When
the time-reversal symmetry is preserved, these mass operators are
all forbidden in the Lagrangian. But four-fermion interaction
terms are still allowed. However, because the fully dressed gauge
field propagators of $a_\mu$ and $b_\mu$ both acquire a $1/k^2$
suppression, when $k$ is large enough, these four-fermion
operators are always irrelevant. Thus at least for large enough
$k$, Eq.~\eqref{qed1} describes a stable $(2+1)d$ CFT.

In Eq.~\eqref{qed1}, a $2\pi-$flux of $a_\mu$ carries charge $n_A$
of the external gauge field $A_\mu$. In the dual theory, the
smallest local gauge invariant operator that carries charge under
$A_\mu$ is (schematically) $(\chi_1)^k \chi_2^\dagger $. This
operator should also have space-time derivatives due to the fermi
statistics of $\chi_1$. This operator carries charge $2n_A$, and
hence can be viewed as the double-monopole operator of $a_\mu$. It
also has a power-law correlation at this CFT, but its scaling
dimension is proportional to $k$ with large $k$.

\subsection{The case with $k=1$}

As was first discussed in Ref.~\onlinecite{senthilfisher}, the
case $k=1$ of Eq.~\eqref{qed1} has an enlarged O(4) global
symmetry, if we ignore the external gauge fields. This O(4)
symmetry becomes very natural knowing the self-duality of the
theory. First of all, when $k=1$, each side of the duality has a
manifest SU(2) global flavor
symmetry~\cite{xudual,karchtong,seiberg2}, thus the symmetry of
the system is at least SO(4)$\sim$SU(2)$\times$SU(2). The $Z_2$
duality transformation is a symmetry that exchanges the two SU(2)
groups, hence the full symmetry of the CFT (assuming $k=1$ is
indeed a stable CFT, as suggested by Ref.~\onlinecite{qedcft}) is
O(4)$=$SO(4)$\times Z_2$, and time-reversal $\mathcal{T}$.

A mass term $\sum_{j = 1}^2 m \bar{\psi}_j\psi_j$ not only breaks
$\mathcal{T}$, but also breaks the $Z_2$ duality symmetry, $i.e.$
it breaks the O(4) symmetry down to SO(4). This is because if we
couple the system to two external SU(2) gauge fields, the SO(4)
invariant mass term will generate different Chern-Simons terms for
the two external SU(2) gauge fields, so it breaks the equivalence
between the two SU(2) symmetries. The existence of an O(4)
breaking but SO(4) invariant mass term implies that, the QED with
$N=2$ is a CFT without any O(4) invariant relevant perturbation,
but once one breaks the O(4) down to SO(4), there will be a
relevant perturbation (if we do not assume an extra
$\mathcal{T}$). This is a signature of this CFT that one can look
for with various numerical methods, and it is fundamentally
different from the ordinary O(4) Wilson-Fisher fixed
point~\footnote{The existence of a relevant SO(4) invariant
deformation of this CFT was first pointed out to us by T. Senthil,
during a private discussion. Here we identify this perturbation as
the mass of the Dirac fermions.}. At the ordinary $3d$ O(4)
Wilson-Fisher fixed point, there is one relevant O(4) invariant
perturbation, but weakly breaking the O(4) down to SO(4) does not
generate any new relevant perturbation.

The effect of the mass term $\sum_{j = 1}^2 m \bar{\psi}_j\psi_j$
can be inferred from Ref.~\onlinecite{xudual}.
Ref.~\onlinecite{xudual} constructed Eq.~\eqref{qed1} with $k=1$ on the
boundary of a $3d$ system, and showed that the mass term $\sum_{j
= 1}^2 m \bar{\psi}_j\psi_j$ generates level $+1$ and $-1$
Chern-Simons terms for the two U(1) external gauge fields, where
the two U(1) symmetries are subgroups of the SU(2) global
symmetries. Formally the level$\pm 1$ U(1) Chern-Simons terms
correspond to level$\pm1/2$ Chern-Simons terms of the SU(2) gauge
fields, where the half-integer level is the sign of anomaly of the
boundary of a $3d$ SPT state, and the anomaly can be cancelled by
the bulk $\Theta-$term. If this theory is realized in a pure $2d$
system, then the external SU(2) gauge fields must receive at least
another level$\pm 1/2$ Chern-Simons terms to cancel the anomaly.
This is the physical meaning of the ``counterterms" introduced in
Ref.~\onlinecite{seiberg2}. For example, if we consider a thin
film of the $3d$ system constructed in Ref.~\onlinecite{xudual},
then this theory on one boundary can receive another level$\pm
1/2$ SU(2) Chern-Simons terms (or level$\pm 1$ U(1) Chern-Simons
terms) from the opposite boundary.

\subsection{Entanglement Entropy}

The entanglement entropy of a $(2+1)d$ CFT across a circle of
radius $R$ takes the general form \beqn S = \alpha
\frac{R}{\epsilon} - F, \eeqn where $\epsilon$ is the
short-distance cutoff. The radius-independent universal term $F$
must decrease under the renormalization group flow~\cite{Ftheorem}
(the F-theorem). Ref.~\onlinecite{pufuentropy} computed $F$ for a
series of $(2+1)d$ CFT with a large flavor number of matter
fields, as $F$ is related to the correctly regularized free energy
of the CFT on a three dimensional sphere~\cite{freeenergy}.

The F-theorem provides us a lower bound for the universal
entanglement entropy $F$ for the CFT in Eq.~\eqref{qed1}. As will
be discussed in section III, a Dirac fermion mass term, which is a
relevant perturbation according to Eq.~\eqref{scaling}, will drive
the CFT into a gapped topological order with $(k^2+1)/2$ different
Abelian anyons (our physical construction in section III
guarantees that $k$ be an odd integer, thus $(k^2+1)/2$ is also an
odd integer). Such Abelian topological order will have a
topological entanglement entropy $F_{\mathrm{topo}} =
\frac{1}{2}\log(\frac{k^2+1}{2})$~\cite{kitaeventropy,wenentropy}.
This implies that $F$ of the CFT in Eq.~\eqref{qed1} must satisfy
\beqn F \geq \frac{1}{2} \log\left(\frac{k^2+1}{2}\right). \eeqn
It is also possible to find an upper bound for $F$, which requires
identifying a UV fixed point that can flow to Eq.~\eqref{qed1}
through a relevant perturbation. Such candidate UV fixed point
could be a supersymmetric version of Eq.~\eqref{qed1} (which was
the strategy used in Ref.~\onlinecite{tarunentropy}). This
supersymmetric CFT presumably flows to Eq.~\eqref{qed1} through a
supersymmetry breaking perturbation.

When $k=1$, a different lower bound of $F$ can be found. In
section III we will see the mass term $m \bar{\psi}_1 \psi_1$ of
fermion $\psi_1$ will drive the system into another CFT that is
conjectured to be dual to the $3d$ XY Wilson-Fisher fixed
point~\cite{wufisher}. If this conjecture is correct, then it
implies that $F_{k = 1}$ must be no smaller than that of the $3d$
XY Wilson-Fisher fixed point.

A direct calculation of $F$ seems to be more involved than the
case studied in Ref.~\onlinecite{pufuentropy}. The reason is that,
after formally integrating out the fermions, the gauge field
$a_\mu$ will acquire an effective Lagrangian, which can be
expanded as a polynomial of $a_\mu$, whose schematic form is
$L_{eff} \sim \sum_n c_n (a_\mu)^n$, and $c_n \sim k^n$ with large
$k$. In Ref.~\onlinecite{pufuentropy}, one can simply keep the
quadratic term of the polynomial as the leading order
approximation, because higher order terms are suppressed in the
large$-N$ limit, since the gauge field fluctuation is at the order
of $O(1/N)$. But in our case, all the higher terms in the
polynomial will be at the same order in the large$-k$ limit, thus
one cannot compute $F$ by truncating the polynomial.

\section{Physical Realization}

\subsection{Bulk Construction}

The theory Eq.~\eqref{qed1} could have various physical
realizations. Here we will construct a $(3+1)d$ gapped bulk state
whose boundary is described by Eq.~\eqref{qed1} with odd integer
$k$. This construction of the bulk state follows similar steps as
Ref.~\onlinecite{maxashvin,xudual}.

(1) We start with a U(1) spin liquid in the $3d$ bulk with a
gapless photon $a_\mu$ and fermionic parton $\psi_\alpha$. Under
time-reversal symmetry, $\psi_\alpha$ transforms as $\mathcal{T}:
\psi \rightarrow i\sigma^y \psi^\dagger$, namely the symmetry group of $\psi$ is
$U(1)_g \times \mathcal{T}$, where $U(1)_g$ is the gauge
symmetry. The electric and magnetic field of $a_\mu$ are odd and
even under time-reversal respectively, which is opposite to the
transformation of the external U(1) gauge field $A_\mu$.

(2) In order to construct Eq.~\eqref{qed1} as the boundary theory,
we assume that the fermion $\psi$ and the $k-$body bound state of
$\psi$ with odd integer $k$ (we denote this bound state as
$\psi^k$) both form a topological insulator in the AIII class, at
the mean field level, $i.e.$ when the gauge field fluctuations are
ignored. Thus at the mean field level, the most natural boundary
state of the system is two flavors of fermions with gauge charge
$1$ and $k$ respectively. A topological $\Theta$-term for $a_\mu$
is generated if we integrate out the gapped partons, with the
theta angle $\Theta = (k^2+1)\pi$. Due to the Witten's effect of
the topological insulator~\cite{qi2008}, a $2\pi$-monopole of
$a_\mu$ will carry a polarization gauge charge
$\frac{\Theta}{2\pi}= (k^2+1)/2$, which is an odd integer, as long
as $k$ is odd. In general, a dyonic excitation in this spin liquid
can be labelled as $(q, m)$ where $q$ is the total gauge charge
and $m$ the monopole number. The Witten's effect then implies $q =
n + \left( \frac{k^2+1}{2} \right) m$, where $n\in\mathbb{Z}$ is
the number of partons $\psi$ attached to the dyon. Thus a
$2\pi-$monopole can be neutralized by binding with $(k^2+1)/2$
{\it holes} of $\psi$. We label this neutralized monopole as the
$(0,1)$ monopole.

(3) In the bulk we condense the bound state of a $(0,1)$ monopole
and a physical boson that carries no gauge charge but one global
U(1) charge~\footnote{We notice that the physical boson must be a
Kramers doublet under $\mathcal{T}$}, $i.e.$ it carries charge
$+1$ under the external gauge field $A_\mu$. This entire bound
state is a gauge neutral, time-reversal invariant, and charge-1
boson. Following the notation in Ref.~\onlinecite{maxashvin}, we
can label all excitations in terms of their quantum numbers $(q,
m, Q, M)$, where $q$ is the gauge charge under $a_\mu$, $m$ is the
monopole number of $a_\mu$, $Q$ is the global U(1) symmetry
charge, and $M$ is the monopole number of the external U(1) gauge
field $A_\mu$. Under this notation, the condensed bound state has
quantum number $(0, 1, 1, 0)$.

(4) The condensate of the aforementioned bound state $(0, 1, 1,
0)$ will confine all the excitations that have nontrivial
statistics with it, including the $\psi$ fermions. But this
condensate does not break any global symmetry. The global U(1)
symmetry is still preserved because the condensed bound state is
also coupled to the dynamical gauge field $a_\mu$. The condensate
does not have any gapless Goldstone mode, which would be a
signature of spontaneous continuous symmetry breaking. If we move
a $2\pi$ Dirac monopole of $A_\mu$ into the bulk, to avoid
confinement caused by the condensate of the $(0, 1, 1, 0)$ bound
state, this $A_\mu$ monopole will automatically pair with a
fermion $\psi$ to form a bound state with quantum number $(1, 0,
0, 1)$, so it has trivial mutual statistics with the condensed
$(0, 1, 1, 0)$ bound state. This deconfined Dirac monopole $(1,
0, 0, 1)$ is neutral under the global U(1) symmetry, but it is a
fermion. This neutral fermionic Dirac monopole of the external
gauge field $A_\mu$ is the characteristic statistical Witten's
effect of the bosonic SPT state with U(1) and time-reversal
symmetry discussed in Ref.~\onlinecite{maxfisher}.

Following Ref.~\onlinecite{maxashvin,xudual}, we can derive the
surface theory of the system. We will skip most of the details
since they are straightforward generalizations of those in
Ref.~\onlinecite{maxashvin,xudual}, and just mention that since we
condense the bound state $(0,1,1,0)$ in the bulk, the surface must
have a $\frac{i}{2\pi}\epsilon_{\mu\nu\lambda}a_\mu \partial_\nu
A_\lambda$ Chern-Simons term, $i.e.$ $n_A=1$.

Our bulk construction indicates that $\psi_1$ can be viewed as a
$k-$body bound state of $\psi_2$. Thus a highly irrelevant
interaction term $\sim \psi_1^\dagger \psi_2^k $ is allowed in
Eq.~\eqref{qed1} (this term must also have space-time derivatives
due to fermion statistics, which renders this term even more
irrelevant). %This irrelevant interaction term will be ignored for%
%the discussion in this paper.
Thus in the IR limit the CFT described in Eq.~\eqref{qed1} has an
emergent global U(1) symmetry in addition to the conservation of
$a_\mu$ flux: $\psi_1$ and $\psi_2$ are conserved separately. The
emergent U(1) symmetry in this construction can also be made an
explicit U(1) symmetry on the lattice, if we assume $\psi_1$ and
$\psi_2$ are two separately conserved fermions. Also, a $2\pi$
monopole of $a_\mu$ carries half-integer polarized fermion number
of $\psi_1$ and $\psi_2$ respectively, thus the flux number of
$b_\mu$ and $c_\mu$ in Eq.~\eqref{qed3} (which are the dual gauge
fields of fermion currents of $\psi_1$ and $\psi_2$ respectively)
must sum to be a multiple of $4\pi$, while each can be a multiple
of $2\pi$. These conditions guarantee that Eq.~\eqref{qed3} is
gauge invariant, and $a_\mu$, $b_\mu$ in Eq.~\eqref{qed1} and
Eq.~\eqref{qed2} both allow $2\pi$ fluxes.

The construction for Eq.~\eqref{qed1} with even integer $k$ is
more involved. We need to assume the existence of another gauge
neutral fermion $\Psi$, in addition to the gauged fermion $\psi$
in the $3d$ bulk. We assume $\psi$, and the bound state of
$\psi^k$ and $\Psi$ both form an AIII topological insulator. The
$2\pi-$monopole of $a_\mu$ carries half-integer polarization gauge
charge. Then in order to confine the gauge field $a_\mu$ in the
bulk, we need to condense the bound state constructed with a
gauge-neutralized $4\pi-$monopole of $a_\mu$ and a physical
electron. Now the boundary of the system is described by
Eq.~\eqref{qed1} with even integer $k$ and $n_A = 1/2$. Thus just
like the case discussed in Ref.~\onlinecite{maxashvin}, this
theory only allows for fluxes of $a_\mu$ that are multiples of
$4\pi$.

\subsection{Anomalous Fractional Quantum Hall states}

%{\it --- Anomalous Fractional Quantum Hall states}

The mass terms for the fermions are allowed once the time-reversal
symmetry is broken, which drives Eq.~\eqref{qed1} and its dual
Eq.~\eqref{qed2} into topological orders with a nonzero Hall
conductivity. There appears to be four different gauge invariant
mass operators: $m_1 \bar{\psi}_1 \psi_1$, $m_2 \bar{\psi}_2
\psi_2$, $m^\prime_1 \bar{\chi}_1 \chi_1$, $m_2^\prime
\bar{\chi}_2 \chi_2$, but they are not independent from each
other. After the masses are turned on, in general the system is
driven into a $\mathcal{T}$-symmetry-breaking fractional quantum
Hall (FQH) phase.

We first compute the Hall conductance of this gapped surface
state. When $m_1 > 0$, $m_2 > 0$, formally the response of the
surface state to the external $A_\mu$ field is described by the
following field theory: \beqn \mathcal{L} = \frac{k^2 + 1}{2}
\frac{i}{4\pi} \epsilon_{\mu\nu\rho}a_\mu
\partial_\nu a_\rho + \frac{i}{2\pi} \epsilon_{\mu\nu\rho}a_\mu
\partial_\nu A_\rho. \label{cs1}\eeqn In the dual side, when
$m_1^\prime < 0$, $m_2^\prime < 0$, the system is described by
\beqn \mathcal{L} &=& - \frac{k^2+1}{2} \frac{i}{4\pi}
\epsilon_{\mu\nu\rho}b_\mu
\partial_\nu b_\rho \cr\cr &-& \frac{ki}{2\pi}
\epsilon_{\mu\nu\rho}b_\mu
\partial_\nu A_\rho - \frac{2i}{4\pi} \epsilon_{\mu\nu\rho}A_\mu
\partial_\nu A_\rho. \label{cs2}\eeqn
Both Eq.~\eqref{cs1} and Eq.~\eqref{cs2} give the same Hall
conductivity \beqn \sigma_H = \frac{2}{k^2+1}. \label{hall1} \eeqn

When $m_1 > 0$, $m_2 < 0$ (or equivalently $m_1^\prime < 0$ and
$m_2^\prime > 0$), the response is also described by a CS theory
similar to Eq.~\eqref{cs1} and Eq.~\eqref{cs2}: \beqn \mathcal{L}
= \frac{k^2 - 1}{2} \frac{i}{4\pi} \epsilon_{\mu\nu\rho}a_\mu
\partial_\nu a_\rho + \frac{i}{2\pi} \epsilon_{\mu\nu\rho}a_\mu
\partial_\nu A_\rho. \label{cs3}
\eeqn
The Hall conductivity of
this state is \beqn \sigma_H = \frac{2}{k^2 - 1}. \label{hall2}
\eeqn

We notice that when $k = 1$, the CS term of $a_\mu$ in Eq. \eqref{cs3}
vanishes, and this phase with $m_1>0, m_2<0$
become a superfluid phase with spontaneous U(1) symmetry breaking,
or equivalently a gapless photon phase in the dual picture. In
this case, if we fix $m_1$ positive, and change the sign of $m_2$,
this boundary system goes through a transition from a quantum Hall
state with $\sigma_H = 1$ to a superfluid phase. This transition
was conjectured to be dual to a $3d$ XY
transition~\cite{wufisher}, which as we discussed in the previous
section, has given us a lower bound of $F$ for the CFT with $k=1$.

These fractional quantum Hall states are anomalous, in the sense
that they cannot be realized in a pure $2d$ system with bosons. To
see why this is the case, we need to learn more about the
topological order of these anomalous FQH states. The most
important fact one needs to keep in mind is that the fundamental
matter field that couples to $a_\mu$ is a fermion. The
Chern-Simons term in Eq.~\eqref{cs1} and \eqref{cs2} attaches a
$\frac{4\pi}{k^2\pm 1}$ $a_\mu$-flux to the fermion, so the
topological twist (or the self statistics) of this excitation is
$\tau = e^{i \theta}=-e^{\frac{2\pi i}{k^2\pm 1}}$. From the
standard flux attachment picture, we also know that the excitation
carries $\frac{2}{k^2\pm 1}$ charge of the global $\mathrm{U}(1)$
symmetry. Other anyonic excitations of this FQH state can be
constructed by fusing multiples of this fundamental anyons
together, and there are in total $\frac{k^2\pm 1}{2}$ of them that
form a $\mathbb{Z}_{(k^2\pm 1)/2}$ fusion group. We will label the
anyon types by an integer $[n]$ (defined mod $\frac{k^2\pm
1}{2}$), with the topological twist and the fractional
$\mathrm{U}(1)$ charge given by
\begin{equation}
\tau_{[n]} = e^{i\theta_{[n]}} = (-1)^{n}e^{\frac{2\pi
in^2}{k^2\pm 1}}, \ \ \ q_{[n]}=\frac{2n}{k^2\pm 1}.
\label{eqn:top_data}
\end{equation}
The statistics of these Abelian anyons are consistent with the
$2d$ topological order described by the $\mathrm{SU}(\frac{k^2\pm
1}{2})_{-1}$ Chern-Simons theory.

Let us start from the state with masses $m_1>0, m_2>0$. To
understand the anomaly on the surface, we imagine adiabatically
inserting a $\Phi=(k^2+1)\pi=\frac{k^2+1}{2}\cdot 2\pi$ flux of
the external U(1) gauge field $A_\mu$ into the surface. Since
$\Phi$ is an integer multiple of $2\pi$, the adiabatic flux
insertion should create a local excitation of the system. Based on
the Hall conductance $\sigma_H=\frac{2}{k^2+1}$, such a flux binds
$Q = 1$ charge of $A_\mu$, and therefore has a self statistics
$e^{i \Phi Q/2}=(-1)^{\frac{k^2+1}{2}}$. For odd $k$,
$\frac{k^2+1}{2}$ is also an odd integer. So inserting $\Phi$ flux
creates a fermionic excitation. On the other hand, the full
braiding of this excitation with an anyonic excitation with
$\mathrm{U}(1)$ charge $q_{[n]} = \frac{2n}{k^2+1}$ in the system
is just given by the Aharonov-Bohm phase $e^{i \Phi q_{[n]}}=1$.
So what we have obtained is a fermionic excitation that has
trivial braiding statistics with all other anyons, which is
impossible in a $2d$ system of bosons. But on the surface of a
$3d$ system, this inconsistency is precisely circumvented by the
statistical Witten effect: insertion of a $2\pi$ flux of $A_\mu$
can be thought of as passing a charge-neutral $2\pi$-monopole of
$A_\mu$ through the surface, which leaves behind an extra fermion
on the surface~\cite{maxfisher}.

This argument of anomaly does not lead to any direct inconsistency
for the $m_1>0, m_2<0$ case since $\frac{k^2-1}{2}$ is an even
integer, so a more elaborate argument is demanded. %We now present
%a general argument establishing a relation between the Hall
%conductance and the fractional $\mathrm{U}(1)$ charges of anyons
%in a topologically ordered phase.
As we have explained, adiabatically inserting a $2\pi$ flux of
$A_\mu$ creates an excitation of the system. We denote the anyon
type of this excitation by $[v]$. One can generalize the previous
argument to show that in a pure two-dimensional bosonic system,
the following two relations must hold~\cite{cheng15,goldhaber}:
\begin{equation}
M_{v,j}=e^{2\pi i q_{[j]}}, \label{M}
\end{equation}
\begin{equation}
\tau_v = e^{i \theta_v}=e^{\pi i\sigma_H}. \label{eqn:hall_v}
\end{equation}
where $j$ denotes the anyon type $[j]$.
$M_{v,j}$ is the mutual braiding statistics between $v$ and $[j]$,
and $\theta_v$ is the self-statistics angle of $v$. These
relations allow one to unambiguously determine $v$ in a $2d$
bosonic system.

Now coming back to the surface FQH state. If we assume that this
surface FQH state can be realized in a pure $2d$ bosonic system,
then based on the data given in Eq.~\eqref{eqn:top_data} and the
general relation Eq.~\eqref{M}, we can determine $v=[1]$. Then
Eq.~\eqref{eqn:hall_v} would imply that $\sigma_H=\frac{2}{k^2\pm
1}+1 + 2\mathbb{Z}$, which differ from Eq.~\eqref{hall1} and
Eq.~\eqref{hall2} by exactly $1$ (mod $2\mathbb{Z}$). This
difference/anomaly can be amended by the $\Theta-$term response in
the bulk with $\Theta = 2\pi$~\cite{senthilashvin}. Or
equivalently, based on the Hall conductivity Eq.~\eqref{hall1} and
Eq.~\eqref{hall2}, the $v$ anyon we would derive in a pure $2d$
bosonic system differs from the anyon type [1] in our surface FQH
state by one local fermion that comes from the statistical
Witten's effect in the $3d$ bulk.

Besides the statistical Witten's effect, some of the boundary FQH states are
anomalous in a different way. The FQH state with $m_1, m_2>0$ is
nonchiral because the Hall conductance of $A_\mu$ and $B_\mu$ in
Eq.~\ref{qed1} are opposite. If we create a domain wall between
two regions with $m_1, m_2>0$ and $m_1, m_2 < 0$, at the domain
wall the chiral central charge is zero. However, if we want to
realize a topological order in $2d$ with the same fusion rule and
statistics as the anyons of the boundary FQH state with $m_1, m_2
>0$, this $2d$ topological order is described by a SU$\left( \frac{k^2 + 1}{2}
\right)_{-1}$ Cherns-Simons field theory, whose boundary has chiral
central charge $c= -\frac{k^2-1}{2}$ mod $8$, and one can readily
check that $ c = 0 \ (\mathrm{mod} \ 8)$ for $k = \pm 1 \
(\mathrm{mod} \ 8)$, and $c = 4 \ (\mathrm{mod} \ 8)$ for $k = \pm
3 \ (\mathrm{mod } \ 8)$. One can drive a purely $2d$ boundary
phase transition to increase the chiral central charge $c$ by a
multiple of $8$ without changing the topological order, due to the
existence of a bosonic state in $2d$ without any topological order
(the so called $E_8$ state~\cite{kitaev_talk, luashvin}), but to
change $c$ by 4 one needs to drive a $3d$ bulk transition, or
attaching the bulk to another bosonic SPT state with time-reversal
symmetry whose boundary is ``half" of the $E_8$
state~\cite{senthilashvin}.

Similarly, when $m_1 > 0$ and $m_2 < 0$, one finds that the
surface state actually has chiral central charge $c = 1$
(assuming $k \neq 1$), but the $2d$ realization of these anyon
types is the SU$\left( \frac{k^2 - 1}{2} \right)_{-1}$ Cherns-Simons
field theory, which has $c = 1 - \frac{k^2-1}{2}$. The
difference between them is again $\frac{k^2-1}{2}$ mod $8$. So the
states with $k = \pm 3 \ (\mathrm{mod } \ 8)$ have another anomaly
independent from the statistical Witten's effect, which can be
amended by a bulk gravitational $\Theta-$term. A similar anomaly
related to the chiral central charge mismatch was discussed
previously in Ref.~\onlinecite{wangcluston}.

The fact that the anomalous FQH state with $m_1 > 0$ and $m_2 < 0$
has chiral central charge $c = 1$, can be understood as the
following. We first create a thin film of the system, so the
entire system is a true $2d$ state without anomaly. On the top
surface, we create a domain wall with $m_1 > 0$, $m_2 < 0$ on the
left, and $m_1 < 0$, $m_2 > 0$ on the right; On the bottom
surface, we create a domain wall with $m_1, m_2 < 0$ on the left,
and $m_1, m_2 > 0$ on the right. Then for the entire $2d$ thin
film, the left side is described by the CS field theory
$\mathrm{SU}(\frac{k^2-1}{2})_{-1} \times
\mathrm{SU}(\frac{k^2+1}{2})_{+1}$, and the right side is
described by the CS field theory
$\mathrm{SU}(\frac{k^2-1}{2})_{+1} \times
\mathrm{SU}(\frac{k^2+1}{2})_{-1}$. The chiral central charge on
the domain wall is $2$, which means that the left side contributes
a chiral central charge $c = 1$. This chiral central charge can
only come from the anomalous FQH state on the top surface, because
from the previous discussion we know that the anomalous FQH state
on the bottom surface is nonchiral.

%If we fix $m_1$ positive, and change the sign of $m_2$, this
%boundary system goes through a transition from a level $(k^2+1)/2$
%CS theory to a $(k^2-1)/2$ CS theory. At the transition, the
%critical point is described by the following Lagrangian: \beqn
%\mathcal{L} = \bar{\psi} \gamma_\mu (\partial_\mu - i a_\mu) \psi
%+ \frac{k^2}{2} \frac{i}{4\pi} \epsilon_{\mu\nu\rho}a_\mu
%\partial_\nu a_\rho; \label{transition1} \eeqn If we fix $m_2$ positive, and change the sign of
%$m_1$, the transition is described by \beqn \mathcal{L} =
%\bar{\psi} \gamma_\mu (\partial_\mu - i k a_\mu) \psi +
%\frac{1}{2} \frac{i}{4\pi} \epsilon_{\mu\nu\rho}a_\mu
%\partial_\nu a_\rho. \label{transition2} \eeqn The new fermion-fermion
%duality implies that Eq.~\ref{transition1} and
%Eq.~\ref{transition2} are dual to each other, hence they have the
%same critical exponent. With large$-k$, their critical exponents
%can be calculated with a $1/k$ expansion with the Lagrangian in
%Eq.~\ref{transition1}.

\section{Summary}

In this paper we studied a series of self-dual $(2+1)d$ CFTs
parameterized by an odd integer $k$. These CFTs are stable for large
enough $k$, $i.e.$ there is no relevant perturbation as long as
certain symmetries (such as U(1), time-reversal, etc.) are
imposed. Unlike the usual cases in $(2+1)d$, the stability of
these CFTs under study do not rely on supersymmetry or a large
number of matter fields. Some questions remain open, and deserve
further study. For example, a direct calculation of the
entanglement entropy $F$ is still demanded. It would also be
interesting to search for other anomalous topological orders
adjacent to the CFT under study.

C. Xu is supported by the David and Lucile Packard Foundation and
NSF Grant No. DMR-1151208. The authors thank T. Senthil and Chong
Wang for very helpful discussions.

\bibliography{QED}

\end{document}